\documentclass[journal]{IEEEtran}

\usepackage{microtype}
\usepackage{graphicx}
\usepackage{subfigure}
\usepackage{booktabs}
\usepackage{hyperref}
\usepackage{amsmath}
\usepackage{multirow}
\usepackage{amsthm}
\usepackage{bbm}
\usepackage{amssymb}
\usepackage{xcolor}
\usepackage{xspace}
\usepackage{bm}
\usepackage{algorithm}
\usepackage{algorithmic}
\usepackage{hyperref}
\usepackage{algorithm}
\usepackage{algorithmic}

\theoremstyle{definition}
\newtheorem{definition}{Definition}

\DeclareMathOperator{\ent}{H}
\DeclareMathOperator{\mi}{I}
\DeclareMathOperator{\prob}{P}
\DeclareMathOperator{\avg}{E}
\DeclareMathOperator{\kl}{\text{D}_\text{KL}}

\begin{document}

\title{InfoShape: Task-Based Neural Data Shaping via Mutual Information}

\author{
    Homa~Esfahanizadeh, William~Wu, Manya~Ghobadi, Regina Barzilay, and Muriel~M\'{e}dard\\
    EECS Department, Massachusetts Institute of Technology (MIT), Cambridge, MA 02139}

\maketitle

\begin{abstract}
The use of mutual information as a tool in private data sharing has remained an open challenge due to the difficulty of its estimation in practice. In this paper, we propose \textit{InfoShape}, a task-based encoder that aims to remove unnecessary sensitive information from training data while maintaining enough relevant information for a particular ML training task. We achieve this goal by utilizing mutual information estimators that are based on neural networks, in order to measure two performance metrics, privacy and utility. Using these together in a Lagrangian optimization, we train a separate neural network as a lossy encoder. We empirically show that \textit{InfoShape} is capable of shaping the encoded samples to be informative for a specific downstream task while eliminating unnecessary sensitive information. Moreover, we demonstrate that the classification accuracy of downstream models has a meaningful connection with our utility and privacy measures.\vspace{-0.1cm}
\end{abstract}

\section{Introduction}
\label{sec:intro}

Mutual information (MI) is a measure to quantify how much information is obtained about one random variable by observing another random variable \cite{Shannon48}. In a data sharing setting, the data-owner often would like to transform their sensitive samples such that only the necessary information for a specific task is preserved, while sensitive information that can be used for adversarial purposes is eliminated. MI is an excellent candidate that can be used to develop task-based compression for data-sharing to address the privacy-utility trade-off problem \cite{Medard:IBprivfunnel}. However, estimating MI without knowing the distribution of original data and transformed data is very difficult, and, consequently, using this critical metric has remained limited. In this paper, we utilize numerical estimation of MI to train a task-based lossy encoder for data sharing.

Machine learning (ML) efforts in various sensitive domains face a major bottleneck due to the shortage of publicly available training data \cite{Ghassemi:mlhealth}. Acquisition and release of sensitive data is a primary issue currently hindering the creation of public large-scale datasets. For example, certain federal regulatory laws such as HIPAA \cite{hipaa} and GDPR \cite{gdpr} prohibit medical centers from sharing their patients' identifiable information. This motivates us to approach the issue from an information theoretic perspective. Our goal is to enable data-owners to eliminate sensitive parts of their data that are not critical for a specific training task before data sharing. We consider a setting where a lossy compressor encodes the data according to two objectives: (i) training a shared model on the combined encoded data of several institutions with a predictive utility that is comparable to the un-encoded baseline; (ii) limiting the use of data for adversarial purposes. In practice, there is a trade-off between the utility and privacy goals.

The state-of-the-art solutions to tackle this privacy-utility trade-off problem mainly involve data-owners sharing their encrypted data, distorted data, or transformed data. Cryptographic methods \cite{Brakerski:efficientfhe,Gentry09,Cheetah} enable training ML models on encrypted data and offer extremely strong security guarantees. However, these methods have a high computational and communication overhead, thereby hindering practical deployment. Distorting the data by adding noise is another solution which can obtain the theoretical notion of differential privacy \cite{Liu:dpimage,Adnan2022,dwork2014algorithmic}, but, unfortunately, often results in notable utility cost. Finally, transformation schemes convert the sensitive data from the original representation to an encoded representation by using a randomly-chosen encoder \cite{Instahide,DAUnTLeSS,Syfer}; however, if the instance of the random encoder chosen by the data-owner is revealed, the original data can be reconstructed.

In contrast, we design an encoding scheme to convert the original representation of the training data into a new representation that excludes sensitive information. Thus, the privacy comes from the \textit{lossy} behaviour of the encoder (i.e., compressor) that we design for a targeted training task. The privacy goal is to limit the disclosed information about sensitive features of a sample given its encoded representation, and the utility goal is to obtain a competent classifier when trained on the encoded data. We propose a dual optimization approach to preserve privacy while maintaining utility. In particular, we use MI to quantify the privacy and utility performance, and we train a neural network that plays the role of our lossy encoder.

There has been recent progress for estimating bounds on the mutual information via numerical methods \cite{Belghazi:mine,Poole:varMI,Song:smile,Choi:remine}. 
We combine the privacy and utility measures using MI estimations into a single loss metric to improve an encoder in its training phase. Once the encoder is trained, it is utilized by individual data owners as a task-based lossy compressor to encode their data for release with the associated labels.

\section{Problem Statement}
\label{sec:setting}
We denote the set of all samples of a distribution by $\mathcal{X}$. Each sample $x \in \mathcal{X}$ is labeled via function $L: \mathcal{X} \rightarrow \mathcal{Y}$. A data-owner has a sensitive dataset $\mathcal{D}\subseteq\mathcal{X}$ that she wishes to outsource to a third party for training a specific classification model (i.e., to learn the function $L$). For privacy concerns, the data-owner first encodes the sensitive data, via an encoder $T:\mathcal{X}\rightarrow \mathcal{Z}$, and then publicly releases the labeled encoded data $\{(T(x),L(x))\}_{x\in\mathcal{D}}$. An adversary may have access to the deposited dataset, but uses it for adversarial purposes, i.e., deriving a sensitive feature $S(x)$ from $T(x)$, where $S:\mathcal{X} \rightarrow \mathcal{Y}'$. We call $L(x)$ and $S(x)$ the public and private label of sample $x\in\mathcal{X}$, respectively. 

The utility goal is to preserve from each sample as much information as needed to train a competitive downstream classification model. The privacy goal is to eliminate unnecessary sensitive data from each sample, which is not critical for the training task but might be misused by an adversary. There are several methods to quantify the privacy and utility performance, and here, we use Shannon entropy \cite{Shannon48}.

\begin{definition} The utility score is negative of the average uncertainty about the public label given its encoded representation,
\begin{equation}\label{eq:utlity}
    M_\text{utility}(T)\triangleq-\ent[L(x)|T(x)].
\end{equation}
\end{definition}
There are two potential ways to express the privacy: Given the encoded representation, one is the average uncertainty about the original sample  and one is the average uncertainty about a sensitive feature of the original sample. While each can be advantageous over the other depending on the problem setting, without loss of generality and for simplicity, we use the second approach in this paper.
\begin{definition} The privacy score is the average uncertainty about the private label given its encoded representation,
\begin{align}
    M_\text{privacy}(T)&\triangleq\ent[S(x)|T(x)].\label{eq:privacy1}
\end{align}
\end{definition}

The privacy and utility are competing targets, and in this paper, we design a lossy encoder that offers a desired trade-off via a Lagrangian optimization. Consider the family of possible encoders as $\mathcal{T}$. An optimal encoder $T^*\in\mathcal{T}$ is obtained as, 
\begin{equation*}
    T^*=\arg\min_{T\in\mathcal{T}}  M_\text{utility}(T)+\lambda M_\text{privacy}(T),
\end{equation*}
{where $\lambda$ is a non-negative metric that controls the trade-off between privacy and utility, chosen to be $1$ in our experiment.}

There has been increasing theoretical interest in using information theoretic measures to encode data for privacy goals. These are organized under the Information Bottleneck method \cite{orig_IB}. However, since it is difficult to calculate these measures due to their dependence on certain (often intractable) probability distributions, they have remained impractical to use. Recent efforts for estimating and incorporating these measures have also faced practical challenges, and deriving connections between optimizing variations of these measures and the success of the task-based encoding (both the utility and privacy aspects) are still open challenges \cite{Belghazi:mine,Song:smile,alemi2017deep}.

\section{Eliminating Sensitive Data}
\label{sec:method}
\begin{figure}
  \centering
  \includegraphics[width=0.38\textwidth]{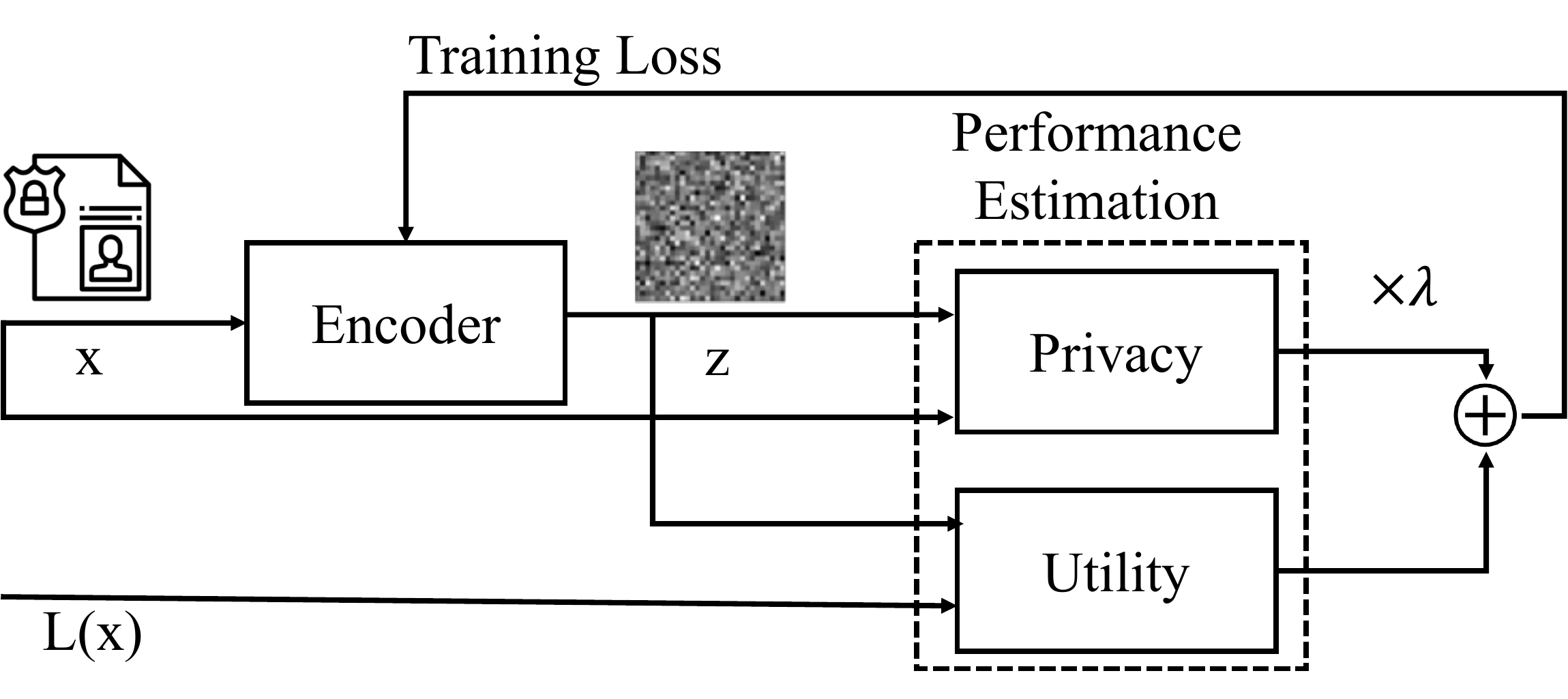}
    \caption{InfoShape design procedure: At each training iteration, the privacy and utility are scored for improving the encoder.}
    \label{fig:TrainingEncoder}
\end{figure}

We propose a dual optimization mechanism, dubbed \textit{InfoShape}, to simultaneously preserve privacy while also maintaining utility on downstream classification tasks, see Fig.~\ref{fig:TrainingEncoder}. We choose the name of \textit{InfoShape} since our scheme trains a neural network encoder to act as a task-specific lossy compressor, by keeping as much relevant information as possible for our intended downstream task while ``shaping'' the data to achieve a private representation.

Consider \textit{InfoShape} as an encoder $T_\theta$ with set of parameters $\theta$ (i.e., an ML model with weights described by $\theta$). We define the loss metric $Q(\theta)$ for evaluating the privacy-utility performance of $T_\theta$ as follows:
\begin{equation}\label{loss}
    Q(\theta)= M_\text{privacy}(T_\theta) + \lambda M_\text{utility}(T_\theta).
\end{equation}
This loss is used for training the encoder by  adjusting $\theta$.

Our optimization problem is to determine the set of parameters $\theta$ such that the loss metric defined in Eq~(\ref{loss}) is minimized. We solve this optimization , i.e., $\theta^*=\arg\min Q(\theta)$, numerically via the stochastic gradient descent (SGD) method \cite{Goodfellow-et-al-2016},
\begin{align}
    &G=\nabla Q(\theta^\textit{itr}),\label{equ:SGD}\;\;\;\;\theta^{\textit{itr}+1}=h\left(\theta^\textit{itr},G\right).
\end{align}
Eq. (\ref{equ:SGD}) shows the gradient of the loss function with respect to weights of the encoder, as well as the weight update step. Here, $h(.)$ is a gradient-based optimizer.

Once the encoder is trained, it can be utilized by individual data-owners as a task-based lossy compressor to encode their data and to enable the release of data for collaborative training.

\subsection{Neural Estimation of Performance Scores}

We utilize neural estimation of MI to numerically approximate the privacy and utility scores. For this, we re-write the privacy and utility scores in Eq.~(\ref{eq:utlity})-(\ref{eq:privacy1}), as follows:
\begin{equation}\label{eq:rewriteMI}
    \begin{split}
        M_\text{utility}(T)&=\mi[L(x);T(x)]-\ent[L(x)],\\
        M_\text{privacy}(T)&=\ent[S(x)]-\mi[S(x);T(x)].
    \end{split}
\end{equation}
Note that the terms $\ent[L(x)]$ and $\ent[S(x)]$ can be computed using the closed-form formulation of Shannon entropy and given the empirical distribution of the public or private labels. Nevertheless, as they 
do not depend on the encoder, they vanish in the gradient.

For training the lossy encoder, we use a set of samples $\{x,L(x),S(x)\}_{x\in\mathcal{P}}$, such that $\mathcal{P}\subset \mathcal{X}\setminus\mathcal{D}$. The underlying distributions are unknown (e.g., $\prob[L(x)|x]$ which characterizes a perfect classifier, and $\prob[S(x)|T(x)]$ which characterizes a computationally unbounded adversary). Consequently, MI is difficult to compute for a finite dataset of high-dimensional inputs \cite{10.1162/089976603321780272}. Thus, we consider tractable variational lower bounds that approximate MI \cite{Belghazi:mine,Choi:remine}. 

Let us consider two random variables $\alpha\in\mathcal{A}$ and $\beta\in\mathcal{B}$. By definition, MI can be expressed in terms of the KL-divergence (a measure of distance between two distributions \cite{Kullback59}) between the joint distribution and multiplications of the marginal distributions of $\alpha$ and $\beta$,
\begin{equation*}
    \mi[\alpha{;}\beta]{=}\kl(\prob[\alpha{,}\beta] {||} {\prob[\alpha]} {\prob[\beta]}){=}{\sum_{\alpha,\beta}}\prob[\alpha,\beta]\log\frac{\prob[\alpha,\beta]}{\prob[\alpha]{}\prob[\beta]}.
\end{equation*}

The Donsker-Varadhan representation of KL-Divergence \cite{Donsker:dvkl} is as follows: Let $\Omega$ be the product sample space of two distributions $\prob_1$ and $\prob_2$.
\begin{equation*}
    \kl(\prob_1 || \prob_2) = \sup_{F: \Omega \rightarrow \mathbb{R}} \avg_{\prob_1}[F] - \log \avg_{\prob_2}[e^F],
\end{equation*}
where the supremum is taken over all functions $F$ such that the two expectations are finite. In \cite{Belghazi:mine}, the sufficiently rich set of functions $F$ is modeled with a neural network $F_\phi$ parameterized with set of weights $\phi\in\Phi$. Let
\begin{equation}\label{ITheta}
    \mi_{\phi}[\alpha;\beta]= \avg_{\prob[\alpha{,}\beta]}[F_{\phi}] - \log \avg_{{\prob[\alpha]} {\prob[\beta]}}[e^{F_{\phi}}].
\end{equation}
The optimal parameter $\phi^*\in\Phi$ that maximizes $\mi_{\phi}[\alpha;\beta]$ can be identified via SGD. $\tilde{\mi}[\alpha;\beta]=\mi_{\phi^*}[\alpha;\beta]$  acts as a lower bound of $\mi[\alpha;\beta]$. In practice, the two expectations in (\ref{ITheta}) are replaced with empirical averages over samples of a minibatch that are drawn according to $\prob[\alpha{,}\beta]$ and ${\prob[\alpha]} {\prob[\beta]}$, respectively. 

Numerically solving (\ref{ITheta}) using SGD to estimate the MI has some practical challenges. In particular, it suffers from bias and a high-variance in the estimation, when MI is large \cite{pmlr-v108-mcallester20a}. To remedy this problem, a regularization term was added to the neural estimation of MI to help with stability \cite{Choi:remine}:
\begin{equation}\label{eq:remine}
\begin{split}
    \mi_{\phi}[\alpha;\beta]&= \avg_{\prob[\alpha{,}\beta]}[F_{\phi}] - \log \avg_{{\prob[\alpha]} {\prob[\beta]}}[e^{F_{\phi}}]\\
    &-0.1(\log \avg_{{\prob[\alpha]} {\prob[\beta]}}[e^{F_{\phi}}])^2.
\end{split}
\end{equation}
The intuition behind the extra regularization term in Eq. (\ref{eq:remine}) is to encourage the optimizer to concentrate on finding one solution in $\{F_\phi:\phi\in\Phi\}$, rather than drifting within a class of equally-behaved functions. To reduce the bias, multiple mini-batches were used to update the MI estimation. For more details we refer the readers to \cite{Choi:remine}, as our core MI estimation method to compute Eq. (\ref{eq:rewriteMI}) in our experimental results.

\subsection{Training Procedure of \textit{InfoShape}} \vspace{0.5cm}

\begin{algorithm}[t]
\caption{Training \textit{InfoShape} (Optimizing $\theta$).}
\label{alg:train}
\begin{algorithmic}[1]
\STATE \textbf{Input:} $\lambda$ and $\{x,L(x),S(x)\}_{x\in\mathcal{P}}$.
\STATE Initialize the encoder parameters $\theta$.
\REPEAT
\STATE Find $\tilde{\mi}[L(x);T_\theta(x)]$ and $\tilde{\mi}[S(x);T_\theta(x)]$.
\STATE Compute $Q(\theta)= M_\text{privacy}(T_\theta) + \lambda M_\text{utility}(T_\theta).$
\STATE Compute $G=\nabla Q(\theta)$ and update $\theta\leftarrow h\left(\theta,G\right)$
\UNTIL{convergence}
\end{algorithmic}
\end{algorithm}
\vspace{-0.2cm}

We present the training procedure for designing \textit{InfoShape} that keeps the necessary information for learning the function $L$, but eliminates the sensitive information needed for learning the function $S$. By training such an encoder, one can ensure that even if an adversary knows the encoder $T_\theta$, they cannot use it to infer sensitive information about the encoded samples. This is because the encoder is not invertible by construction, and even by having $T_\theta$ and public samples with disclosed private labels $\{x,S(x)\}_{x\in\mathcal{P}}$, one cannot train a competent classifier that infers sensitive information in the encoded domain. This fact is also supported by our experimental results showing that training a classifier to estimate the private labels in the encoded domain is highly unsuccessful compared to training a classifier to estimate the public labels.\footnote{Our experiments implemented the task-based encoder using a simple neural network. If an adversary obtains a good prediction accuracy for sensitive features, one can use a more complicated encoder architecture or increase the training iterations to kill more sensitive data.}

Algorithm~\ref{alg:train} shows the step-by-step procedure for training $T_\theta$. The input for the algorithm is the trade-off parameter $\lambda$, and a public set of samples with both public and private labels $\{x,L(x),S(x)\}_{x\in\mathcal{P}}$, line~1. We first sample a set of random layer weights $\theta$ for the encoder, line~2. 
Now, starting from the first iteration and until convergence, we iteratively evaluate the performance of $T_\theta$ and update its weights accordingly, lines~4-6. In particular, we need to estimate ${\mi}[L(x);T_\theta(x)]$ and ${\mi}[S(x);T_\theta(x)]$ (line~4), for which we use the neural MI estimator in \cite[Algorithm~1]{Choi:remine} due to its numerical stability advantages. 
We have provided public access to our code and data at \href{https://github.com/billywu1029/infoshape}{https://github.com/billywu1029/infoshape}.


\section{Simulation Results}
\label{sec:experiments}
In this section, we empirically show the potential of our tasked-based lossy encoding scheme through evaluation of its privacy and utility scores. For this purpose, we consider classifiers that are trained and tested on {four separate datasets: original data, randomly encoded data, noisy data (by adding independent Gaussian noise per sample), and encoded data using \textit{InfoShape}}. For each training dataset, we compare the accuracy between two classifiers: one that identifies the public label (faithful user) and one that identifies the private label (unfaithful user). Further, we show the estimation of privacy score and utility score at different training epochs of our \textit{InfoShape} encoder using the MI estimators.

\subsection{Dataset 1: Synthetic Multi-Class Dataset}

In this subsection, we utilize a balanced dataset which contains 10,000 samples, each being a vector of size 10 features belonging to one of four different classes.\footnote{We used the sklearn.datasets.make{\textunderscore}classification Python library function to generate this random 4-class dataset.} Each sample has 3 primitive features, 2 redundant features, and 5 noisy features.  We created 2 clusters of samples per class, and each cluster is constructed as follows: The primitive features are first drawn independently from a standard Gaussian distribution and then randomly linearly combined within each cluster in order to add covariance. The clusters are placed on the vertices of a 3d hypercube with sides of length 2. For each sample, the redundant features are generated as random linear combinations of the primitive features, and the noisy (useless) features are attained using random noise. Samples and the features are then shuffled, and 99\% of samples within each cluster are assigned to the same class. In order to obtain both private labels and public labels for our dataset, we represent each class with two bits (i.e., $\{00,01,10,11\}$), whose most significant and least significant bits represent the private and public labels, respectively.

We first describe our \textit{InfoShape} design parameters: Our encoder is modeled with a dense neural network with 10 input nodes, one intermediate layer with 10 nodes, 3 output nodes, and Tanh non-linearity. We use 50 epochs for training, Adam optimizer, and learning rate 1e-3. As for the loss function, we need to estimate $\mi[L(x);T(x)]$ and $\mi[S(x);T(x)]$, see Eq.~(\ref{eq:rewriteMI}). We use the procedure in \cite{Choi:remine}, with custom architectures for the neural networks used for MI estimation: a dense neural network with 4 input nodes (3 for the encoded sample and 1 for the label), two intermediate layers with 100 nodes each, 1 output node, and ReLu non-linearity. For the back propagation algorithm to estimate MI, we run 2000 iterations using an Adam optimizer and a learning rate of 1e-4, with a batch size of 2000 for each iteration. To encourage numerical stability, we only update the gradients for the MI estimation after accumulating and averaging gradients for 10 iterations. 

\begin{figure}
    \centering
    \vspace{-0.15cm}\includegraphics[width=0.52\textwidth]{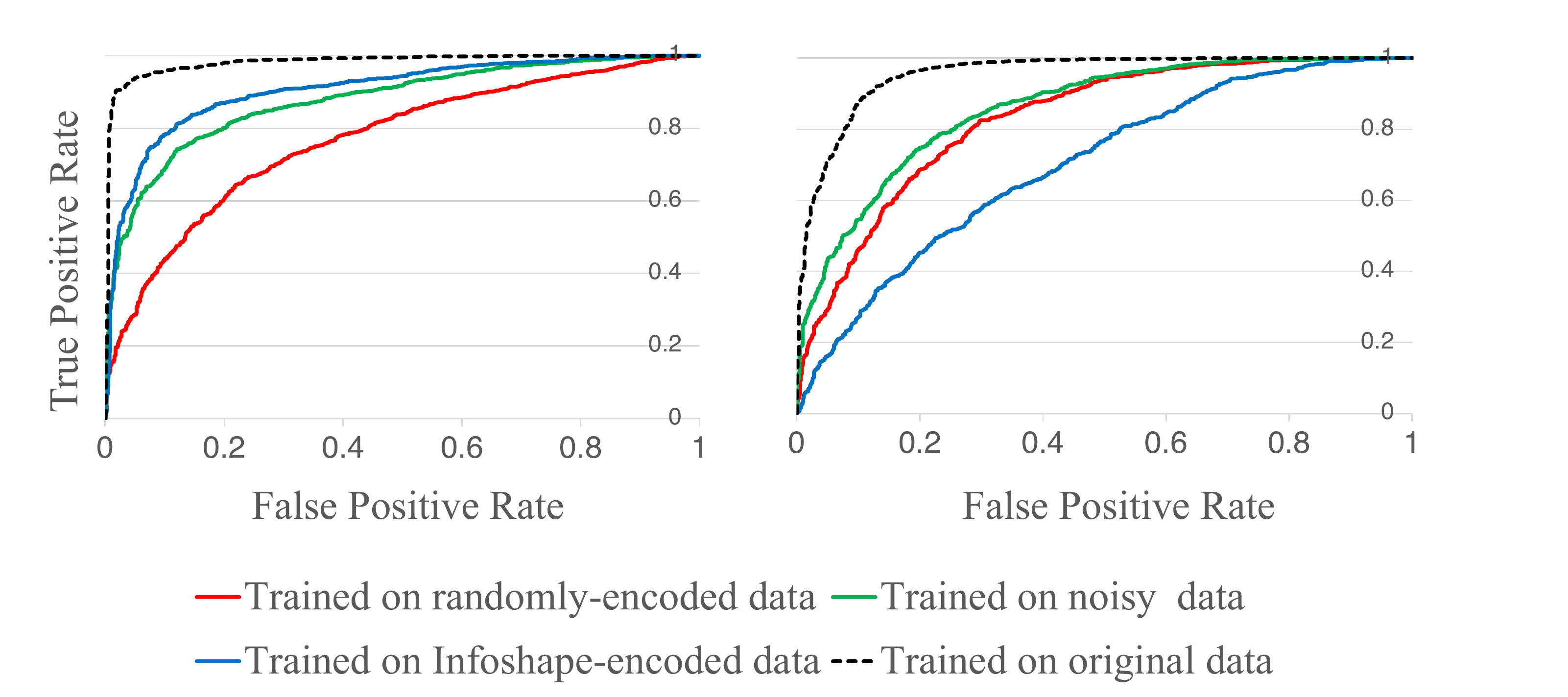}\vspace{-0.3cm}
    \caption{Validation ROC for modeling (left) public labeling function and (right) sensitive labeling function. The ROC AUC has slightly decreased from $0.98$ for original dataset to $0.91$ for \textit{InfoShape}-encoded dataset regarding the public labeling function, while it has dramatically decreased from $0.96$ to $0.66$ regarding the private labeling function.}
    \label{fig:ROC}
\end{figure}

Fig.~\ref{fig:ROC} depicts the validation ROC, a graph showing the performance of a classification model at all classification thresholds: the left and right sub-figures show the classification performance for the public label and private label, respectively. The architecture of the classifiers is a dense neural network with 3 input nodes, one intermediate layer with 20 nodes, 1 output node with a Sigmoid activation, and ReLU non-linearity between other layers. We use SGD optimizer, batch size 100, learning rate 1e-4, and 50 epochs for training the classifiers, and we use 20\% of data for the validation of trained models. 

{The dashed black line in Fig.~\ref{fig:ROC} represents the accuracy of classifiers trained on the original un-encoded data. The red, green, and blue lines show the accuracy of classifiers trained on randomly-encoded data, and noisy data, and \textit{InfoShape}-encoded data respectively. Compared to the random encoder baseline, our proposed scheme enables a higher accuracy for the public label and a lower accuracy for the sensitive label, showing success of the task-specific lossy compression. Compared to the Gaussian noise baseline, \textit{InfoShape}-trained model results in slightly
better utility performance (AUC\footnote{Area under the ROC curve} of $0.91$ vs. $0.88$) and much better privacy preservation (AUC $0.66$ vs. $0.86$), verifying that adding noise indistinguishably
degrades the AUC of both the sensitive and public features. This experiment shows that an adversary who has access to some public relevant data and the exact encoder would not achieve a good accuracy to identify the sensitive features of encoded data.}

\begin{figure}
    \centering
    \includegraphics[width=0.50\textwidth]{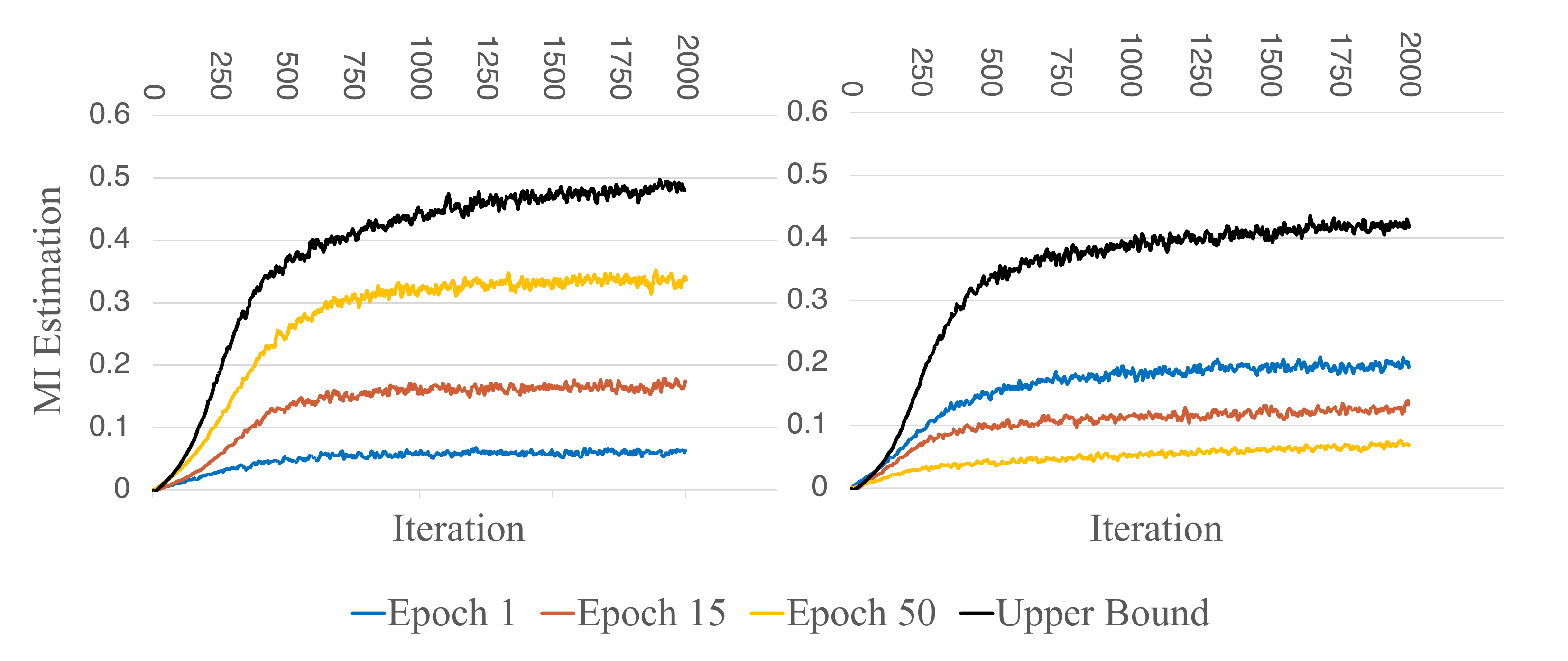}\vspace{-0.3cm}
    \caption{MI estimation between encoded sample and its public label and private label (left $\mi[L(x);T(x)]$; right $\mi[S(x);T(x)]$). Graphs are smoothed with a moving average filter. The curves for $\mi[L(x);x]$ and $\mi[S(x);x]$ are shown in black for reference.}
    \label{fig:MI}
\end{figure}

Fig.~\ref{fig:MI} shows the estimated values of $\mi(L(x),T(x))$ and $\mi(S(x), T(x))$ per iteration obtained by the MI estimators at various epochs of the encoder training. The average value of the final iterations are used in Eq.~(\ref{eq:rewriteMI}) to obtain the loss value at each epoch. The estimators show an increase by a factor of $5.1$ in the MI between encoded data and public label and a decrease by a factor of $0.65$ in the MI between encoded data and sensitive label, between training epochs 1 and 50.

\begin{figure}
    \centering
    \begin{tabular}{cc}
    \hspace{-0.4cm}\includegraphics[width=0.275\textwidth]{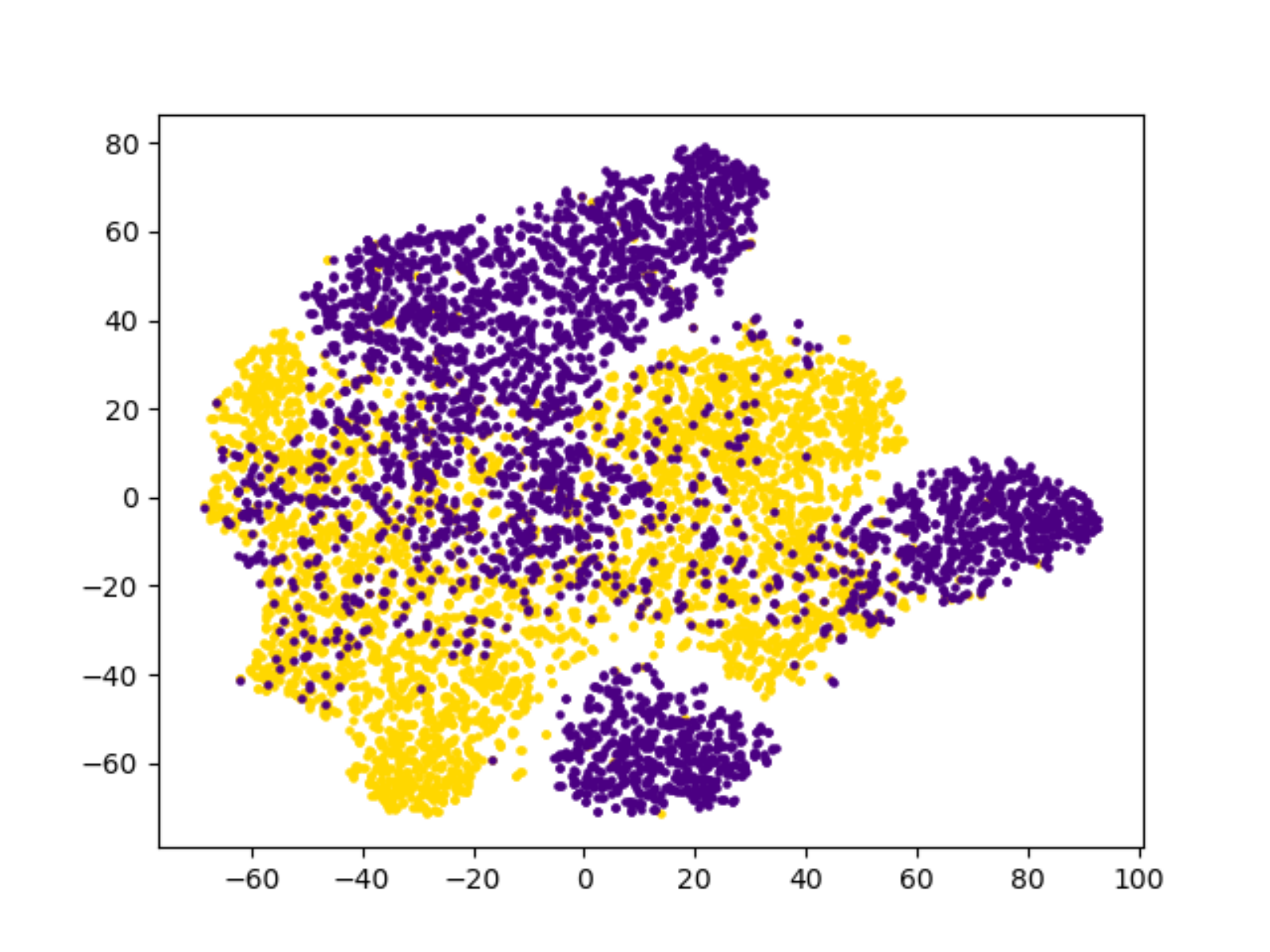}&
    \hspace{-0.9cm}\includegraphics[width=0.275\textwidth]{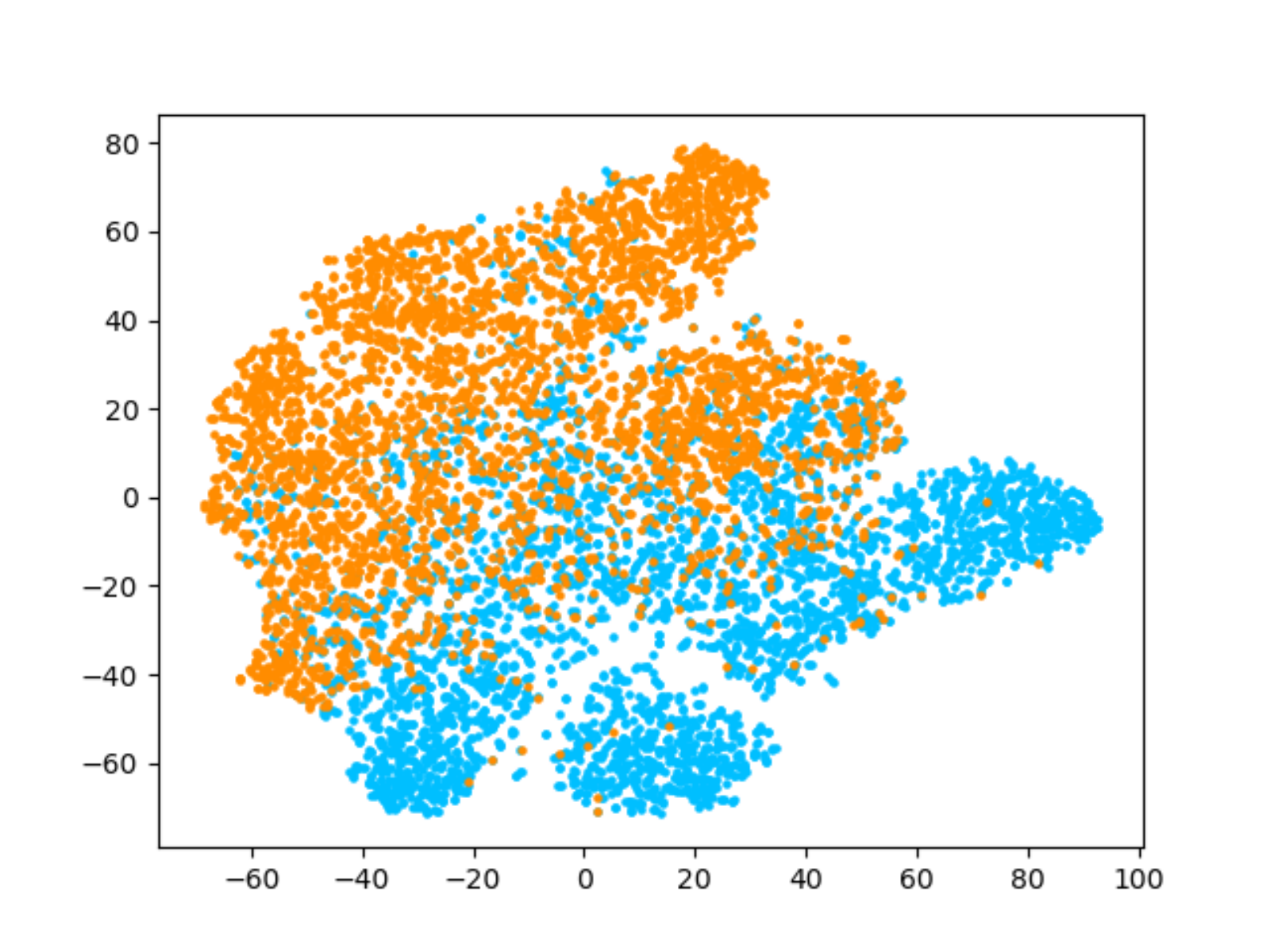}\vspace{-0.5cm}\\
    \hspace{-0.4cm}\includegraphics[width=0.275\textwidth]{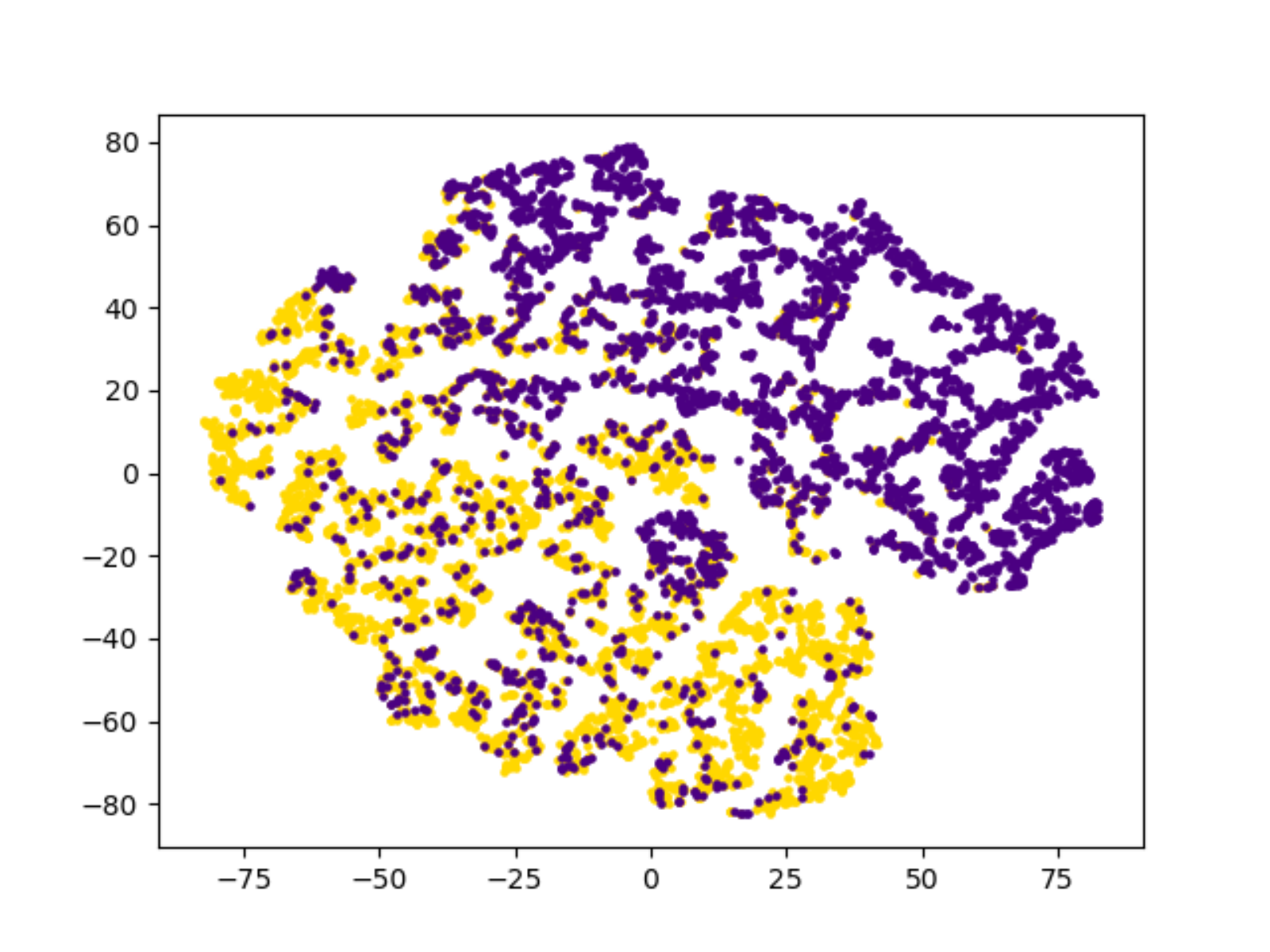}&
    \hspace{-0.9cm}\includegraphics[width=0.275\textwidth]{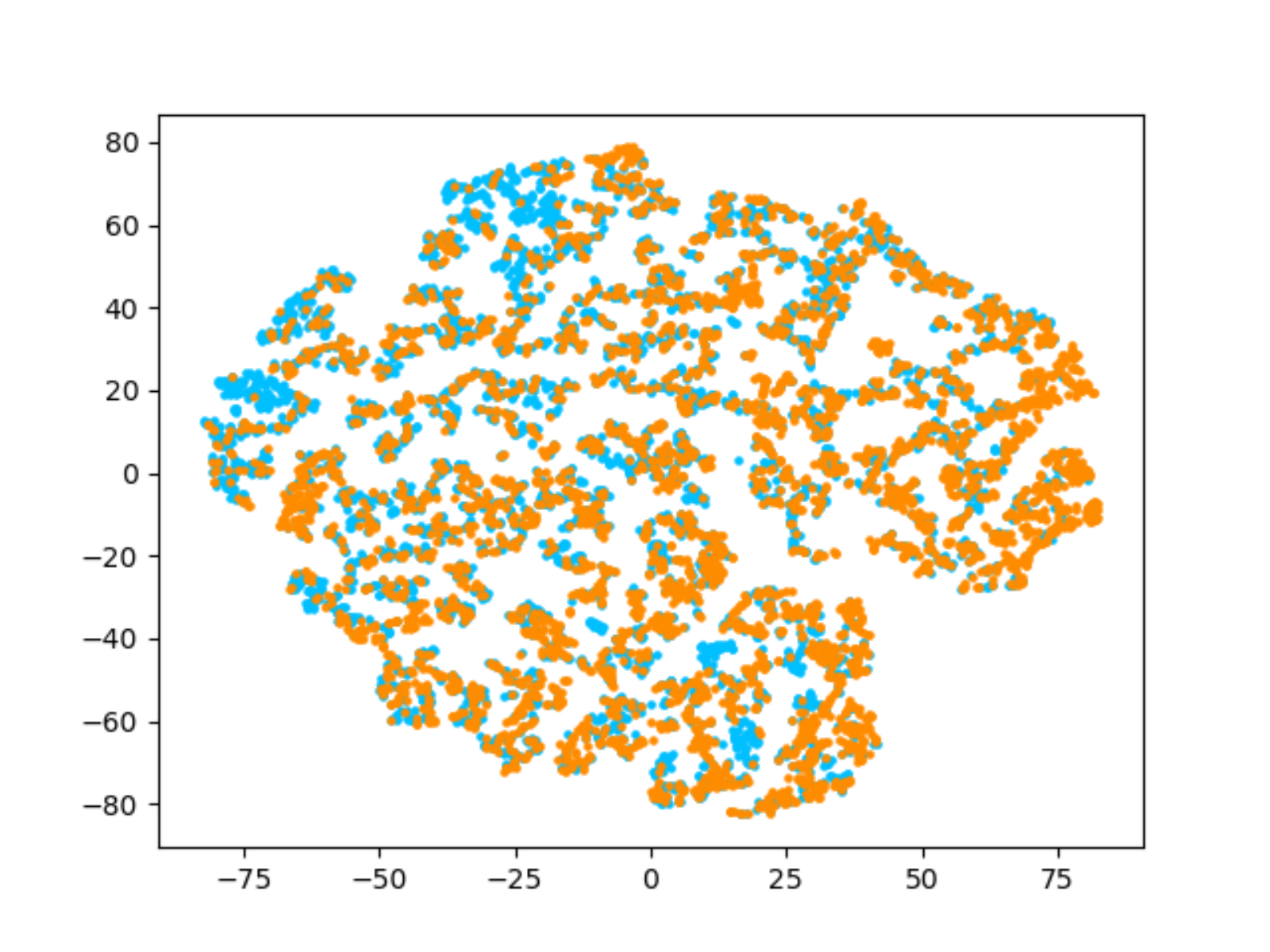}
    \end{tabular}
    \caption{2D visualizations of dataset samples, colored based on their public label (left) and private label (right). The top panel is dedicated to the original samples, and the bottom panel is dedicated to the \textit{InfoShape}-encoded samples.}
    \label{fig:D1:t_SNE}
\end{figure}

Finally, we use the t-distributed stochastic neighbor embedding (t-SNE) \cite{t-SNE} method, for visualization of high-dimensional dataset samples by giving each sample a location in a two-dimensional map. The results are given in Fig.~\ref{fig:D1:t_SNE}, where the top and bottom panels show the visualizations of the original samples and \textit{InfoShape}-encoded samples, respectively. In each panel, samples are colored based on their public label on the left, and based on their private label on the right. These results verify that the task-specific encoder reduce the cluster-ability of samples based on their private label, while preserving the separability of samples based on their public labels.

\subsection{Dataset 2: Digit MNIST Dataset}

\begin{figure}
    \centering
    \includegraphics[width=0.48\textwidth]{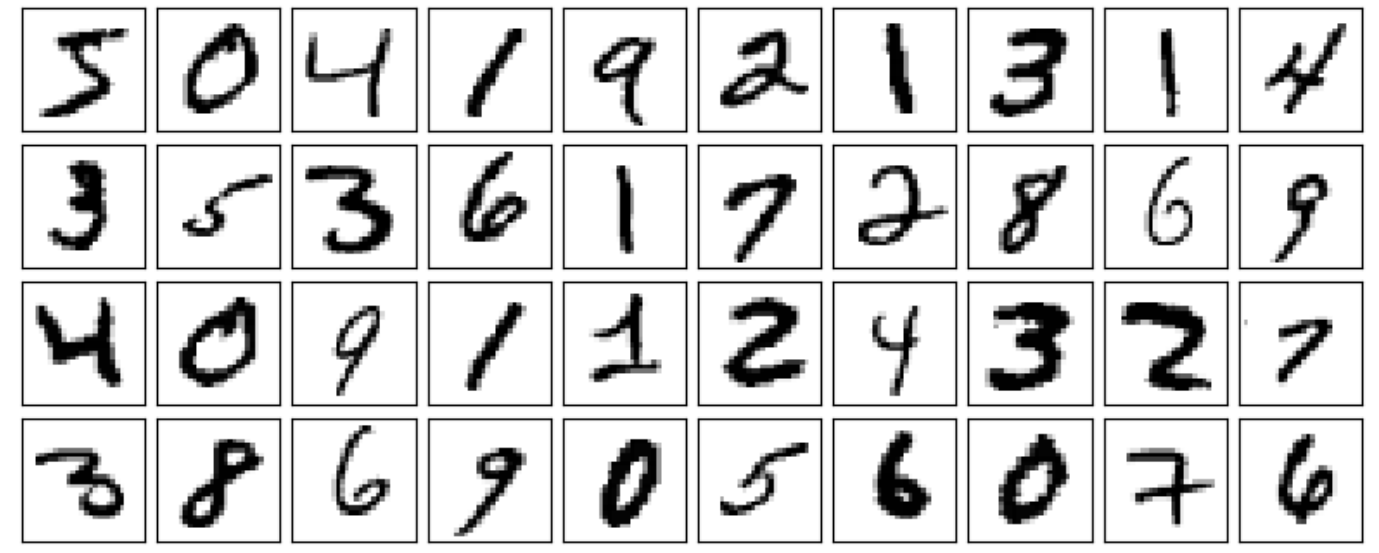}
    \caption{Several randomly-selected samples in MNIST dataset.}
    \label{fig:D2:samples}
\end{figure}

In this subsection, we use MNIST, which is a balanced dataset of handwritten digits \cite{deng2012mnist}. It contains $60,000$ training images and $10,000$ testing images. The samples are $28\times28$ grayscale images, normalized to be within $0$ and $1$. The original labels are integers in $\{0,1,\dots,9\}$, corresponding to the digit illustrated by each hand-written image. Fig.~\ref{fig:D2:samples} shows several arbitrarily-chosen samples of the MNIST dataset. We define the private label as if the digit is even or odd, and the private label as if the digit is greater than $4$ or not. The goal is to encode the dataset such that the encoded samples can be used for training a classifier to identify if an encoded sample is even or odd, while hiding the fact that if the encoded sample is greater than $4$ or not.

Our \textit{InfoShape} encoder is modeled with a dense neural network with 784 input nodes (one input node per pixel), one intermediate layer with 50 nodes, 10 output nodes, and Tanh non-linearity. We use 50 epochs for training, Adam optimizer, and learning rate 1e-3. 

The architecture of the neural network used for MI estimation is a dense neural network with 11 input nodes (10 for the encoded sample and 1 for the label), two intermediate layers with 100 nodes each, 1 output node, and ReLu non-linearity. For the back propagation algorithm to estimate MI, we run 2000 iterations using an Adam optimizer and a learning rate of 1e-4, with a batch size of 2000 for each iteration. To encourage numerical stability, we only update the gradients for the MI estimation after accumulating and averaging gradients for 10 iterations.

\begin{figure}
    \centering
    \hspace{-0.35cm}\includegraphics[width=0.50\textwidth]{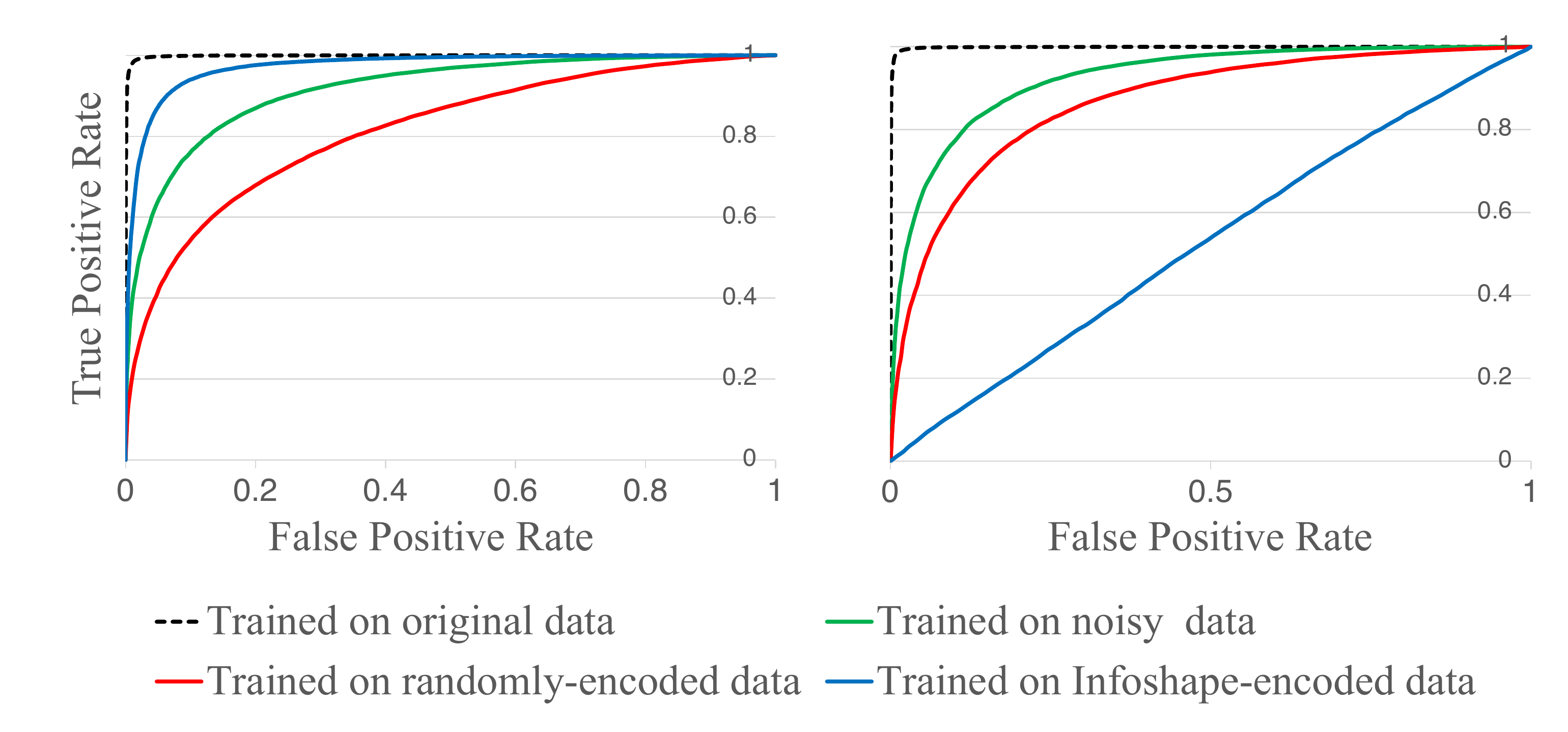}\vspace{-0.3cm}
    \caption{Validation ROC for modeling (left) public labeling function and (right) sensitive labeling function. The ROC AUC has slightly decreased from $0.999$ for original dataset to $0.973$ for \textit{InfoShape}-encoded dataset regarding the public labeling function, while it has dramatically decreased from $0.999$ to $0.527$ regarding the private labeling function.}
    \label{fig:D2:ROC}
\end{figure}

Fig.~\ref{fig:D2:ROC} depicts the validation ROC, where the left and right sub-figures show the classification performance for the public label and private label, respectively. The classifiers are trained on four different datasets, i.e., original, noisy, randomly-encoded, and \textit{InfoShape}-encoded, marked with different colors. The architecture of the classifiers is a dense neural network with 10 input nodes (784 input nodes when trained on original dataset), one intermediate layer with 50 nodes, 1 output node with a Sigmoid activation, and ReLU non-linearity between other layers. We use SGD optimizer, batch size 100, learning rate 1e-4, and 10 epochs for training the classifiers. Compared to the original dataset, our proposed scheme enables a comparable prediction accuracy for the public label and a close-to-random accuracy for the sensitive label, highlighting success of the task-specific lossy compression. The other two baselines result in a higher privacy leakage, while showing a worse predictive utility as well.

\begin{figure}
    \centering
    \includegraphics[width=0.50\textwidth]{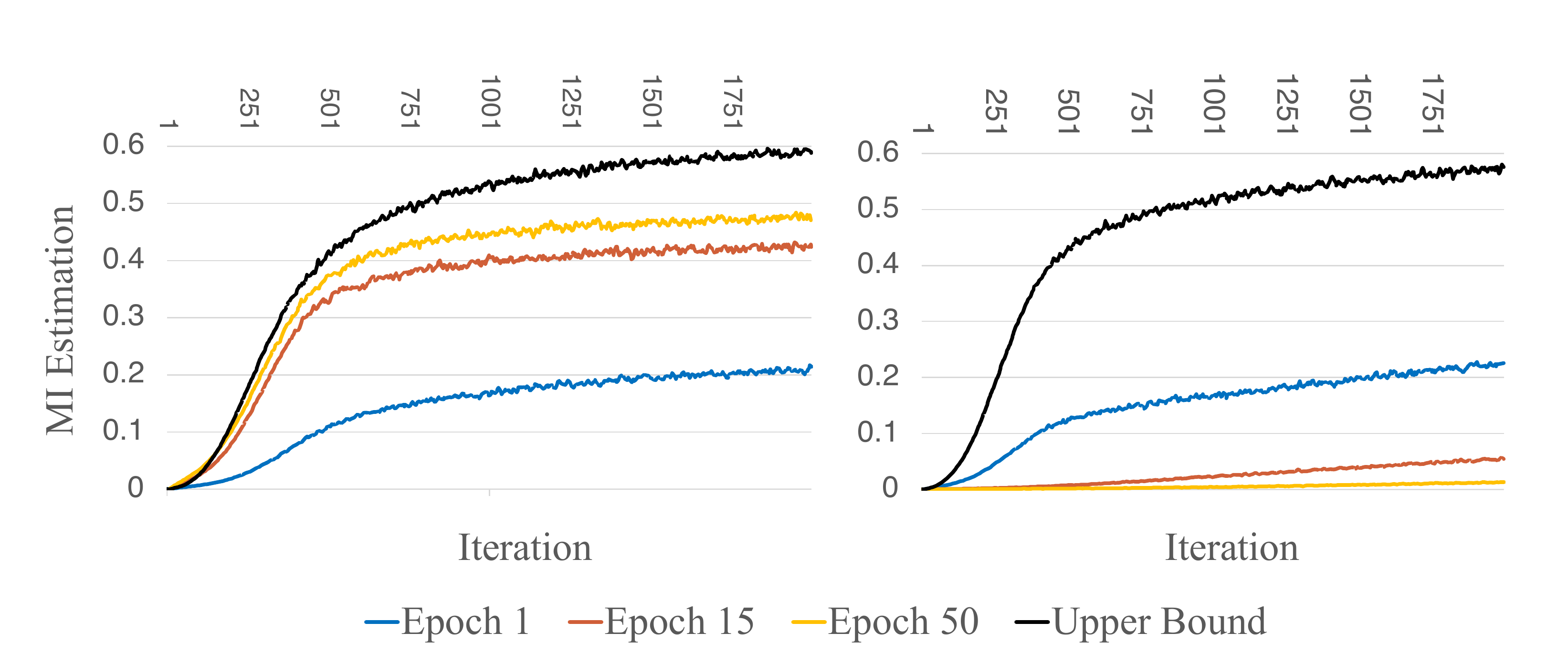}\vspace{-0.3cm}
    \caption{MI estimation between encoded sample and its public label and private label (left $\mi[L(x);T(x)]$; right $\mi[S(x);T(x)]$). Graphs are smoothed with a moving average filter. The curves for $\mi[L(x);x]$ and $\mi[S(x);x]$ are shown in black for reference.}
    \label{fig:D2:MI}
\end{figure}

\begin{figure}
    \centering
    \begin{tabular}{cc}
    \hspace{-0.4cm}\includegraphics[width=0.275\textwidth]{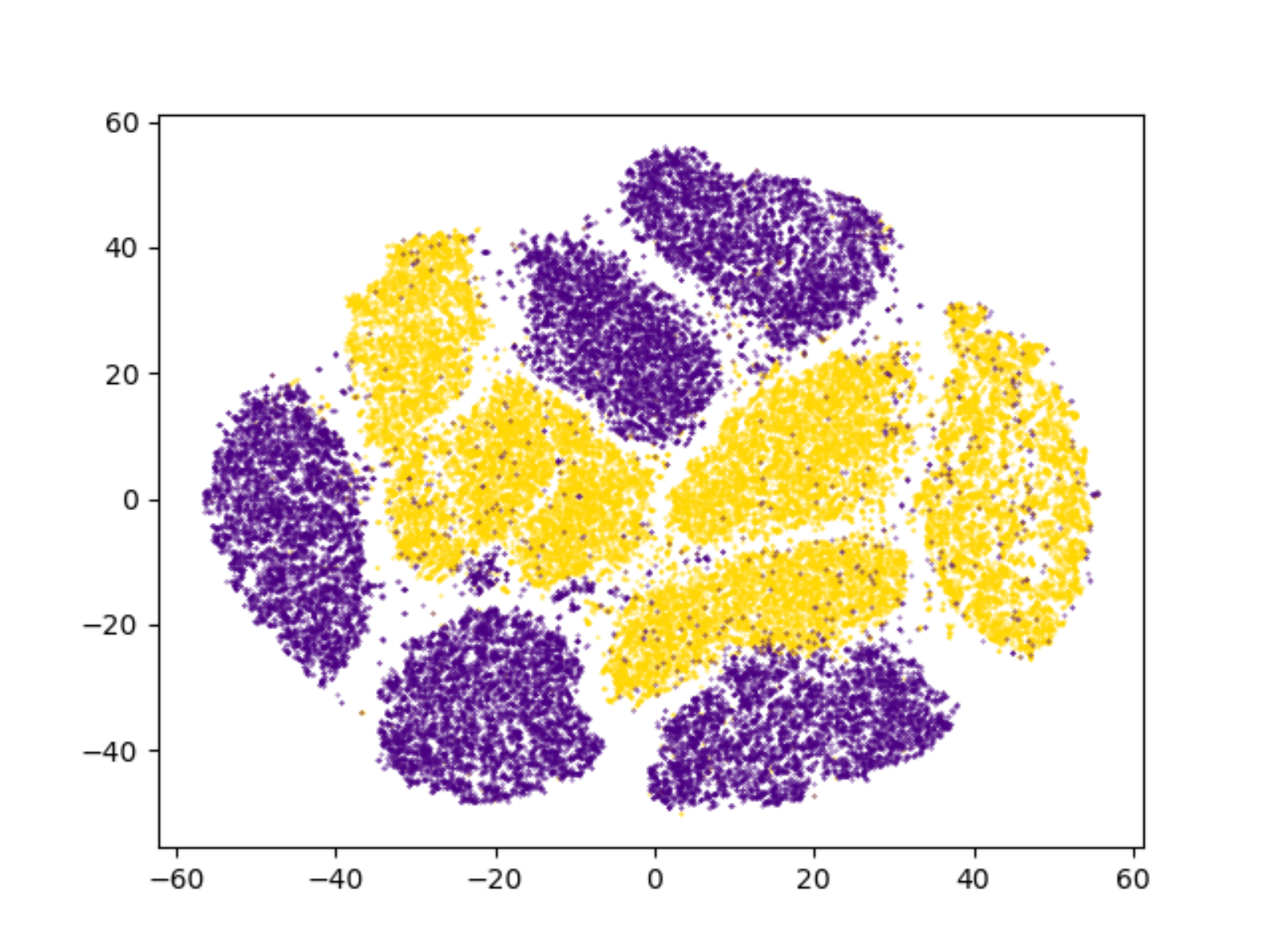}&
    \hspace{-0.9cm}\includegraphics[width=0.275\textwidth]{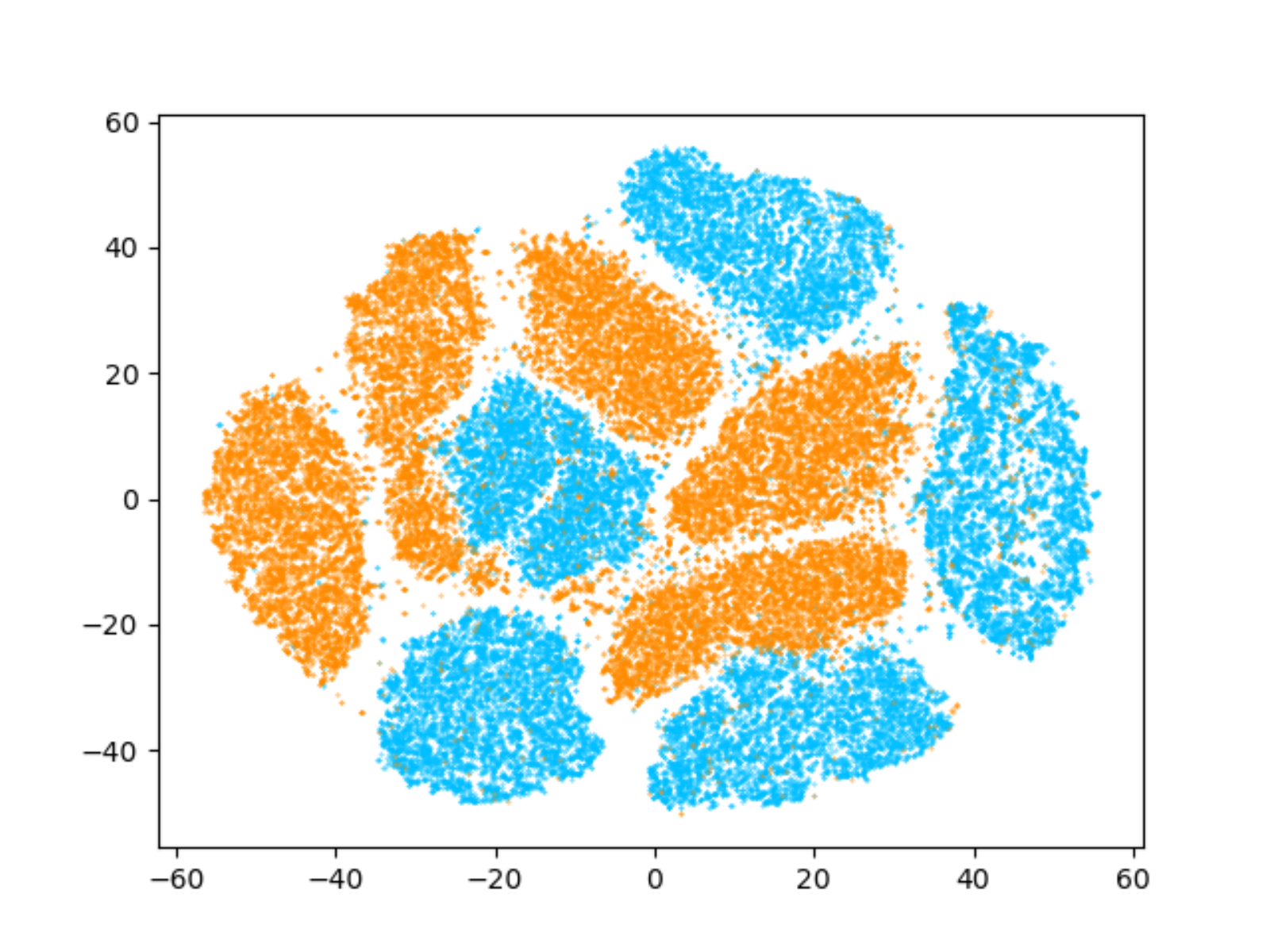}\vspace{-0.5cm}\\
    \hspace{-0.4cm}\includegraphics[width=0.275\textwidth]{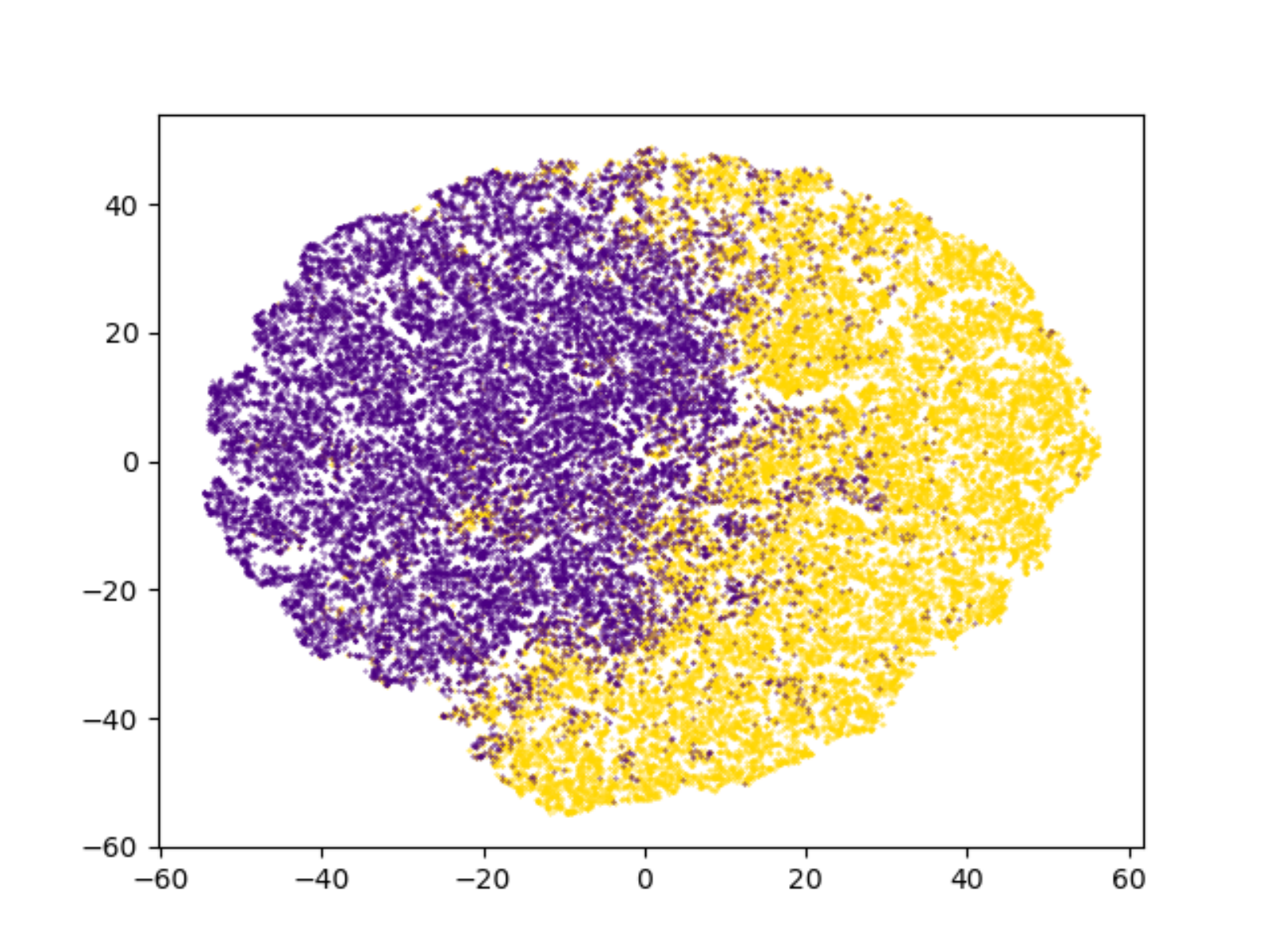}&
    \hspace{-0.9cm}\includegraphics[width=0.275\textwidth]{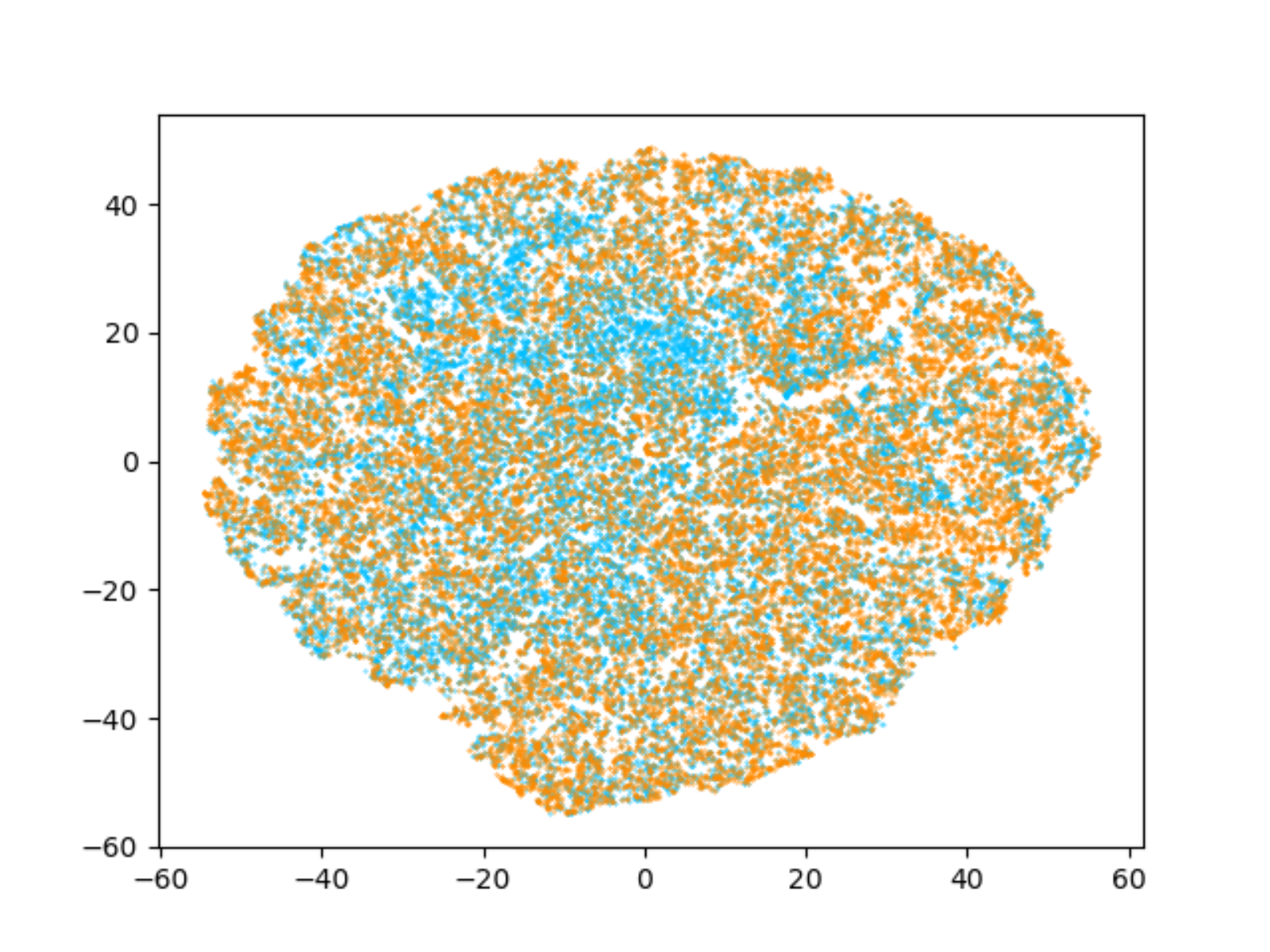}
    \end{tabular}
    \caption{2D visualizations of dataset samples, colored based on their public label (left) and private label (right). The top panel is dedicated to the original samples, and the bottom panel is dedicated to the \textit{InfoShape}-encoded samples.}
    \label{fig:D2:t_SNE}
\end{figure}

Fig.~\ref{fig:D2:MI} shows the estimated values of $\mi(L(x),T(x))$ and $\mi(S(x), T(x))$ per iteration obtained by the MI estimators at various epochs of the encoder training.
Finally, we use the t-SNE method for 2D visualization of high-dimensional dataset samples. The results are given in Fig.~\ref{fig:D2:t_SNE}, verifying that the task-specific encoder makes distinguishing between samples based on their private label visually impossible, while preserving the visual separability between samples based on their public labels.

\section{Conclusion and Future Work}
In this paper, we presented \textit{InfoShape}, a neural network trained using neural MI estimations to tackle the critical privacy-utility trade-off problem in data sharing. 
{The presented framework can be combined with future MI estimators that offer better numerical stability to apply to real world data, e.g., imaging data in healthcare. The usage of other information measures, such as guesswork, and developing mechanisms to add noise just to the sensitive content of samples remain as future work.}

\bibliographystyle{IEEEbib}
\bibliography{references.bib}

\end{document}